\documentclass[pre,aps,reprint]{revtex4-1}
\usepackage[utf8]{inputenc}
\usepackage{amsmath,graphicx,color}
\newcommand{\avg}[1]{\langle #1\rangle}
\newcommand{\vek}[1]{\boldsymbol{#1}}
\begin{document}
\title{Identification and modeling of discoverers in online social systems}
\author{Matúš Medo$^1$, Manuel S. Mariani$^1$, An Zeng$^{1,2}$, Yi-Cheng Zhang$^1$}
\affiliation{$^1$Department of Physics, University of Fribourg, 1700 Fribourg, Switzerland\\
$^2$School of Systems Science, Beijing Normal University, 100875 Beijing, P.R. China}

\begin{abstract}
The dynamics of individuals is of essential importance for understanding the evolution of social systems. Most existing models assume that individuals in diverse systems, ranging from social networks to e-commerce, all tend to what is already popular. We develop an analytical time-aware framework which shows that when individuals make choices---which item to buy, for example---in online social systems, a small fraction of them is consistently successful in discovering popular items long before they actually become popular. We argue that these users, whom we refer to as discoverers, are fundamentally different from the previously known opinion leaders, influentials, and innovators. We use the proposed framework to demonstrate that discoverers are present in a wide range of systems. Once identified, they can be used to predict the future success of items. We propose a network model which reproduces the discovery patterns observed in the real data. Furthermore, data produced by the model pose a fundamental challenge to classical ranking algorithms which neglect the time of link creation and thus fail to discriminate between discoverers and ordinary users in the data. Our results open the door to qualitative and quantitative study of fine temporal patterns in social systems and have far-reaching implications for network modeling and algorithm design.
\end{abstract}

\maketitle

\section{Introduction}
The digital age provides us with unprecedented amounts of information about our society. The collected data are increasingly available at fine temporal resolution which permits to progress from rudimentary mechanisms in complex systems, such as preferential attachment \cite{barabasi1999emergence,mitzenmacher2004brief,barabasi2009scale}, to their refined versions where the fitness of individual nodes and aging play a fundamental role \cite{Medo11,Wang13}. In a similar way, the Poisson assumption of the distribution of human activity in time~\cite{greene1997production} has been replaced with models based on a priority queuing process~\cite{barabasi2005origin} and a cascading non-homogeneous Poisson process~\cite{malmgren2008poissonian} that correctly reproduce the observed bursts of activity~\cite{eckmann2004entropy,malmgren2009universality}. Bigger and better data continue to foster our understanding and modeling of the human behavior.

We focus here on data produced by various online systems where users acquire items: buy products, borrow DVDs, or watch videos, for example. This kind of data is at the center of attention of the recommender systems community which aims at predicting items that an individual user might appreciate~\cite{schafer1999recommender,adomavicius2005toward,koren2009matrix,lu2012recommender}. The user-item data can be represented and modeled by a growing network where users are connected with the collected items \cite{zhou2007bipartite,newman2010networks}. Preferential attachment assumes that the rate at which items attract new connections from users is proportional to the number of connections that items already have~\cite{mitzenmacher2004brief}. Models based on preferential attachment have been applied in a wide range of systems \cite{albert2002statistical}. However, all models to date consider a homogeneous population composed of users driven by item popularity which is modulated by item fitness or aging or both in more elaborate models \cite{dorogovtsev2000evolution,yook2002modeling,Medo11}.

We develop here a statistical framework based on data with time information and a new metric, user surprisal, to show that users in social systems are essentially heterogeneous in their collection patterns: while the majority of users are subject to preferential attachment and usually collect popular items, some users frequently attach to little popular items that at the same time eventually become hugely popular. We focus on the latter group of users and suggest a criterion to select those of them who are statistically significant---we refer them as discoverers here. We use our framework to find discoverers in data from a number of real systems and illustrate that the identified discoverers can be used to predict the future popular items. The success of discoverers cannot be due to the fact that they act as opinion leaders or influentials \cite{katz1955personal,bass1969product,watts2007influentials,cha2010measuring,aral2012identifying} because the possibility for users to directly influence each other is absent in most of the datasets studied here. By contrast to innovators who act in the first stage of innovation diffusion~\cite{rogers2010diffusion}, discoverers are distinguished by consistency with which they achieve discoveries and by not relying on their social status or social contacts. In other words, the discoverers discussed here are a genuinely new component in the much-studied diffusion of innovations~\cite{rogers2010diffusion,bikhchandani1992theory,gomez2013diffusion} and the evolution of complex systems~\cite{bar1997dynamics,albert2002statistical}.

To provide a possible explanation for the observed collection patterns of discoverers, we generalize a recent network growth model~\cite{Medo11,medo2014statistical} by assuming that there are two kinds of users: those who are driven by item popularity and those who are driven by item fitness. Since high fitness items often become very popular (though, similarly as in real systems~\cite{basuroy2003critical,salganik2006experimental}, the correlation is not perfect) and fitness-driven users are often among the first users who collect these items, the model produces similar discovery patterns as those observed in the real data. We provide basic analytical results for the model and study its dependence on model parameters.

Finally, we demonstrate that the artificial data generated by this model contradict the score-feedback mechanism used by many ranking algorithms on networks \cite{franceschet2011pagerank}, of which PageRank\cite{brin1998anatomy,langville2004deeper} and HITS (Hyperlink-Induced Topic Search)~\cite{kleinberg1999authoritative} are the prime examples. We show that these algorithms, which only act on a static snapshot of the system, consequently fail to individuate the fitness-sensitive users in model data. Our surprisal metric, although not devised to rank the users, overcomes the problem by taking the chronology of link creation into account. Our observations point out for the first time that classical ranking algorithms on networks may be inadequate for a broad class of systems. The only way to overcome this inadequacy is to develop new algorithms motivated by and benefiting from temporal patterns in real data.

The paper is organized as follows. In Section~\ref{sec:stat}, we introduce the statistical framework for the identification of discoverers.
In Section~\ref{sec:real}, we apply the framework on datasets from two real online systems and show that they both feature strong support for discoverers. In Section~\ref{sec:model}, we present a network growth model which reproduces the discovery patterns observed in the real data. We also show here that while the classical ranking algorithms fail to discriminate the users in model data, our new framework performs well in this respect.

\section{Statistical procedure}
\label{sec:stat}
We assume that the input data have a bipartite structure where there are $U$ users, $I$ items, and $L$ links that always connect a user and an item. We label the users with Latin letters ($i, j,\dots$) and the items with Greek letters ($\alpha, \beta,\dots$) to make the notation more transparent.

\subsection{Discoveries and user surprisal}
To find the users who act as discoverers of highly popular content, we devise a simple yet effective procedure. We choose a small fraction $f_D$ of the most popular items and track the users who are among the first $N_D$ users connecting with them; here $N_D$ is a small parameter. We label these early links as \emph{discoveries} of the eventually popular content. The number of links created by user $i$ and the number of thus-achieved discoveries are denoted by $k_i$ and $d_i$, respectively.

To evaluate whether a user under- or outperforms in making discoveries, we formulate the null hypothesis $H_0$ that all users are equally likely to make a discovery by each collected item. Denoting the total number of discoveries and links as $D=\sum_i d_i$ and $L=\sum_i k_i$, respectively, the probability of discovery for each individual link under $H_0$ is $p_D(H_0) = D/L$. Under the null hypothesis, discoveries are independent and equally likely---their number for any given user is thus driven by the simple binomial distribution. This allows us to compute the probability that user $i$ makes at least $d_i$ discoveries as
\begin{equation}
\label{NullProb}
P^0(d_i\vert k_i, p_D, H_0) = \sum_{n = d_i}^{k_i} {k_i \choose n} p_D^n (1 - p_D)^{k_i - n}.
\end{equation}
By summing up over $d_i$ discoveries or more, we make sure that the probability $P^0$ can become very small only if the user makes too many discoveries, not too few, in comparison with the user's degree $k_i$. Note that the expected number of discoveries of user $i$ is $\avg{d_i}=p_D k_i$ and the total expected number of discoveries is therefore
\begin{equation}
\sum_i \avg{d_i} = \sum_i \frac DL\, k_i = D.
\end{equation}
The binomial distribution for the number of discoveries by individual users and the real number of discoveries are thus compatible with each other. Note that the null hypothesis effectively decouples the users whose discoveries are assumed to be independent of the discoveries made by the others. While this is not strictly true on a link-by-link basis---a user sometimes creates a link at a moment when there are no discoveries possible---it still holds for each user overall because every user makes several links and, moreover, they are free to choose the time when the links are made.

To quantify the extent to which is the behavior of user $i$ incompatible with the null hypothesis, we introduce user \emph{surprisal} (also referred to as self-information \cite{cover2012elements})
\begin{equation}
\label{surp}
s_i := -\ln P^0(d_i\vert k_i, p_D, H_0).
\end{equation}
The higher the surprisal, the more unlikely the user under $H_0$. The lowest possible surprisal value $S_i=0$ and the highest possible surprisal value $S_i = -k_i\ln p_D$ are achieved when $D_i = 0$ and $D_i = k_i$, respectively. Albeit the proposed procedure to compute user surprisal is not well adapted to the extreme case of a user who collects all items, we can consider it as an instructive example. A user who collects all items as the first one (for example by setting up an automaton that periodically checks the system and collects any new items that appear) naturally makes many discoveries---each link to an items which eventually ends among the $f_DI$ most popular items counts as one discovery. At the same time, the number of discoveries expected under the null hypothesis is $p_DI$. When $p_D > f_D$, the actual number of discoveries does not exceed the expectations under the null hypothesis and the user's surprisal value is thus small. The conclusion is that as long as $p_D$ is greater than $f_D$, even collecting each item as the first one is no guarantee of achieving high surprisal.

\subsection{The bootstrap analysis}
To evaluate whether a user's discovery behavior is compatible with the null hypothesis, we use parametric bootstrap \cite{wasserman2004all}. Using the discovery probability $p_D$, we generate the number of discoveries under $H_0$ for each user according to Eq.~(\ref{NullProb}), compute the corresponding bootstrap surprisal value, and consequently compute the largest bootstrap surprisal value found for any of the users. By repeating this procedure many times (we use 10,000 independent bootstrap realizations), we find the average largest surprisal value in bootstrap $\avg{S_{\mathrm{max}}^B}$. Any user whose real surprisal is higher than this value is referred to as \emph{discoverer}; the number of discoverers is labeled as $U_D$. In addition to this quantity, bootstrap is used to obtain Zipf plots of bootstrap surprisal values which are used for comparison with Zipf plots of real surprisal values (given a set of values, a Zipf plot is a plot of the logarithm of a value rank versus the logarithm of the value itself~\cite{stanley1995zipf}).

\begin{figure*}
\centering
\includegraphics[scale = 0.66]{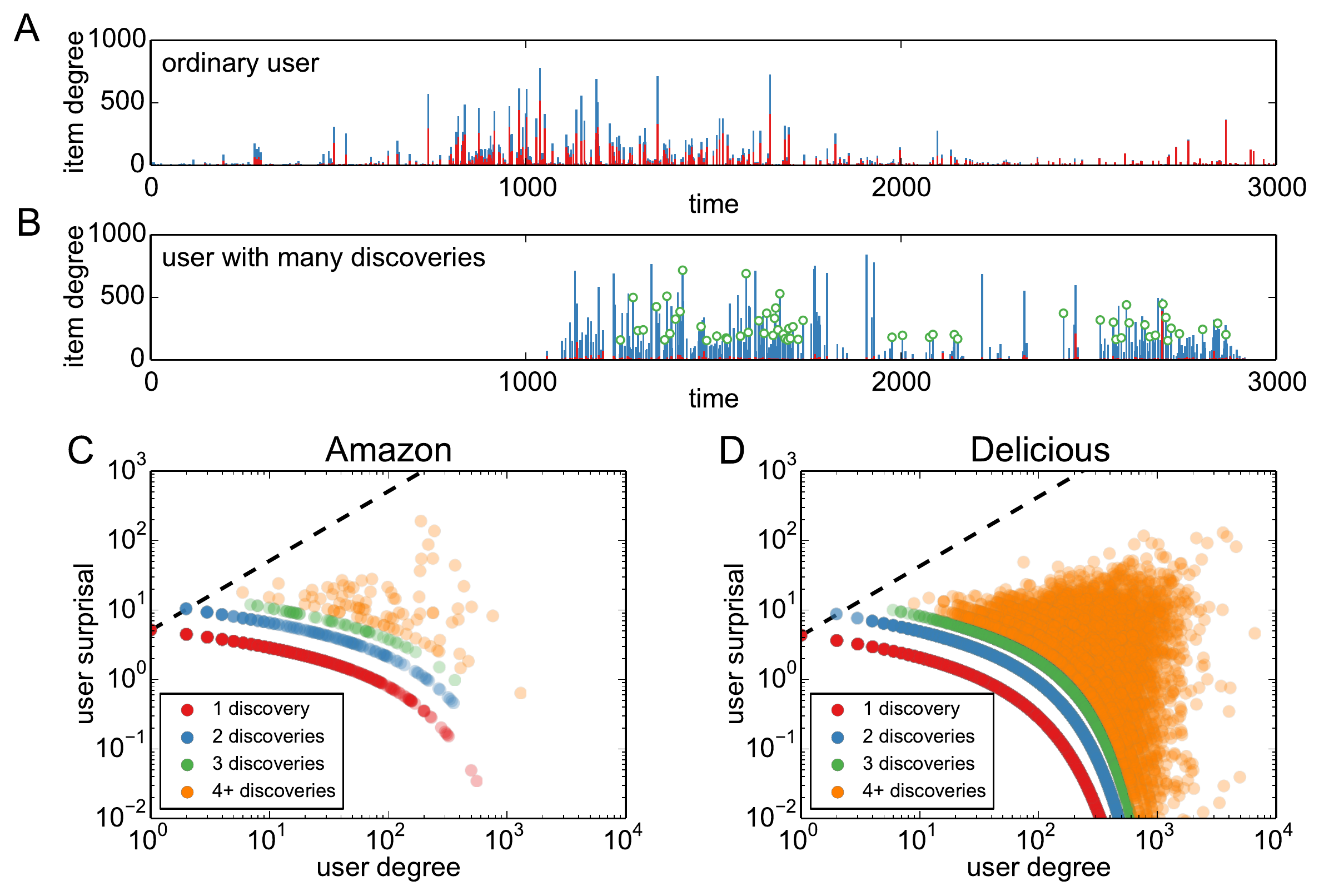}
\caption{Discoveries and user surprisal in real data. (A, B) A comparison of the linking patterns of an ordinary user and a ``discoverer'' in the Amazon data (see Fig. S1 for results on the Delicious data). Each bar here corresponds to a collected object where the red and blue part show item popularity at the time when the user collected it and the final popularity, respectively. Green circles mark those collected items that are eventually identified as discoveries (see the definition in text). (C, D) Scatter plots of user degree and user surprisal in the Amazon and Delicious data. Users are color-coded according to their number of discoveries. The dashed lines mark the maximal achievable user surprisal at a given user degree. All results are for $f_D=1\%$ and $N_D=5$.}
\label{fig:disc_real}
\end{figure*}

In each bootstrap realization, there is some non-zero probability that some users reach higher surprisal than $\avg{S_{\mathrm{max}}^B}$. To find this ``null level'' of discoverers, one can apply the same procedure to bootstrap surprisal values: all users whose bootstrap surprisal is higher than $\avg{S_{\mathrm{max}}^B}$ are classified as discoverers and their number is denoted by $U_D^0$ (the superscript $0$ stands for the null level). Simulations show that $U_D^0$ reaches small values (of the order of one) for the two investigated datasets. We can conclude that this null level of discoverers is of little practical significance and one can omit it in the surprisal analysis and computation of $U_D$.

\section{Real data analysis}
\label{sec:real}
To test the proposed procedure, we use data on DVD purchases at Amazon.com and personal bookmark collections at Delicious.com. See Supplementary Information (SI) for results on four additional datasets.

\subsection{Data description}
Amazon DVD review data were obtained from \url{snap.stanford.edu/data/web-Movies.html} \cite{mcauley2013amateurs}. After data cleaning (merging distinct items which actually correspond to the same product---different releases of a DVD are the typical example of this phenomenon---and removing duplicate reviews), there are 1,901,110 reviews in the integer scale 1--5 from 889,066 users for 141,039 items. While the data span 5,546 days (August 1997-October 2012), we only use the data from days 2,000 to 5,000 because the rest of the data shows comparably low activity of users. To obtain an unweighted bipartite network, we neglect all reviews with rating 3 or less and represent all reviews with rating 4 or 5 as links between the corresponding user and item. After this operation, there are 713,581 links whereas 406,275 users and 76,205 items have at least one link.

Delicious.com is a web site that allows users to store, share, and discover web bookmarks. Delicious bookmark collections were obtained by downloading publicly-available data from the social bookmarking website \url{delicious.com} in May 2008. Due to processing speed constraints, we randomly sampled 50\% of all users available in the source data and included all their bookmarks. To avoid the possible ambiguity of various web addresses pointing to the same web page, reduce the number of items and thus increase the data density, bookmarks are represented only by their base www-address without the initial protocol specification, possible leading ``www.'' and the trailing slash (\emph{e.g.}, \url{http://www.edition.cnn.com/US/} is modified to \url{edition.cnn.com}); each www-address is then represented as an item-node and connected with the users who have collected it. Time stamps are counted in hours from 01/09/2003 and run from 0 to 36,027. For the same user activity reasons as in Amazon, we only use the data from hours 15,000 to 35,000. There are 107,810 users, 2,435,912 items and 9,322,949 links in the resulting data. We have analyzed also data where the full address hierarchy is preserved (\emph{e.g.}, \url{edition.cnn.com/US} instead of the previously mentioned \url{edition.cnn.com}) and found the same behavior as presented here.

\begin{figure*}
\centering
\includegraphics[scale = 0.66]{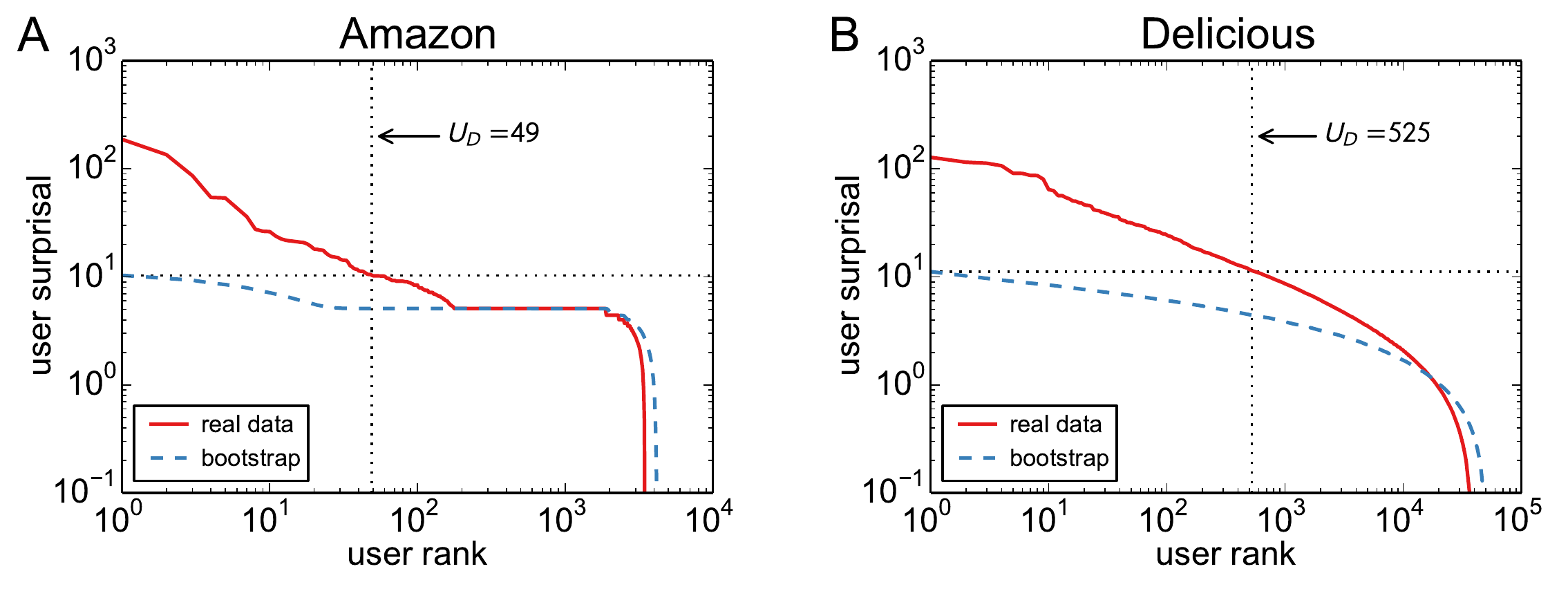}
\caption{Zipf plots of user surprisal in real data and in bootstrap. All results are for $f_D=1\%$ and $N_D=5$.}
\label{fig:bootstrap}
\end{figure*}

\begin{figure*}
\centering
\includegraphics[scale = 0.66]{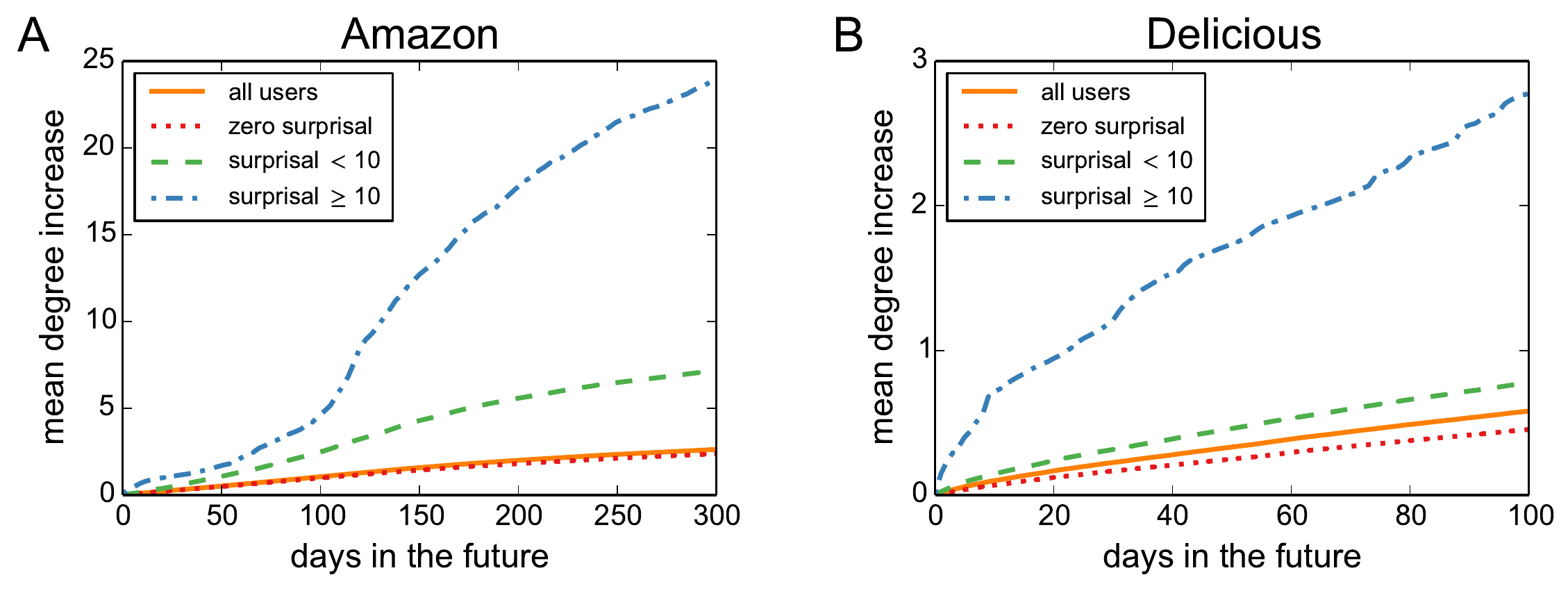}
\caption{Future degree evolution of the target items collected by users of different surprisal in Amazon (A) and Delicious (B) data. The items collected by high surprisal users become significantly more popular (according to the Mann-Whitney test) than target items collected by users of low or zero surprisal. The popularity
ratio between the high and zero surprisal group at the end of the future time window is $10.0$ and $6.1$ for the Amazon and Delicious data, respectively.}
\label{fig:future}
\end{figure*}

\subsection{Results on real data}
Figure \ref{fig:disc_real} shows the discovery patterns and user surprisal in the real datasets. Panels \ref{fig:disc_real}A and \ref{fig:disc_real}B compare the linking patterns of two Amazon users of different surprisal. The ``ordinary user'' either collects popular items late or collects unpopular items and thus achieves no discoveries. By contrast, the ``user with many discoveries'', though only active later during the dataset's timespan, is frequently among the first to collect eventually popular items and achieves 59 discoveries in 283 links whereas the overall discovery probability is $p_D\approx0.5\%$ which for the given number of links corresponds to $1.4$ discoveries on average. Panels \ref{fig:disc_real}C and \ref{fig:disc_real}D further show the degree and surprisal values in the analyzed data. While the maximal possible surprisal value of an individual user grows linearly with user degree (depicted with dashed lines), user activity alone is no guarantee of high surprisal and top surprisal values are achieved by some moderately active users (see Tab.~S2 for the list of users with the highest surprisal values in the two datasets). Finally, one can see here that when the number of discoveries is fixed, the surprisal value decreases with user degree.

Results of the bootstrap analysis in Figure~\ref{fig:bootstrap} show that the largest surprisal values in bootstrap realizations sampled under $H_0$ are never as high as the largest surprisal in real data. For $f_D=1\%$ and $N_D=5$, there are $49$ and $525$ identified discoverers in the Amazon and Delicious data, respectively ($0.01\%$ and $0.49\%$ of all users, respectively). The highest surprisal values correspond to the $P^0$ probabilities $10^{-131}$ and $10^{-56}$ for the Amazon and Delicious data, respectively. The same kind of discovery behavior in four additional data sets is reported in Fig. S2. The SI further demonstrates that there is no particular time bias in the discovery patterns (e.g., discoverers are not those who happen to be active earlier or longer than the others) and the discoveries are made continuously during the system's lifetime (Figures S3 and S4, respectively). While numerical values of surprisal depend on parameters $f_D$ and $N_D$, the resulting ranking of users by their surprisal is rather stable (see Fig. S5). Figure S6 finally demonstrates that the ranking of users by their surprisal does not change considerably when part of the data is taken into account. We can conclude that the null hypothesis of user homogeneity needs to be rejected because some users are indeed significantly more successful than the others in early collecting eventually popular items. This phenomenon is not restricted to particular conditions and emerges consistently in systems where individuals are free to choose among many heterogeneous items.

We next investigate whether the presence of users who make discoveries more often than the others is of some practical significance. To this end, we generate multiple data subsets and in each of them define young items with exactly one link as the target items whose future popularity is to be predicted (see SI for details). Since the information on these items is extremely limited and the social network of users either absent in the studied systems or not known to us, traditional methods for prediction of popularity of online content cannot be used here~\cite{szabo2010predicting,yang2011patterns,cheng2014can}. We divide users in each subset into three groups: zero, low, and high surprisal users (the threshold between low and high surprisal is set to $10$ which is close to the average highest surprisal value in bootstrap in both data sets). The data that come after a given subset are then used to evaluate the future degree evolution for the target items collected by users from different groups. Figure~\ref{fig:future} demonstrates that the target items chosen by users of high surprisal become significantly more popular than those chosen by users of zero or low surprisal. This shows that surprisal not only quantifies users' past behavior but it also has predictive power.

\section{Network model}
\label{sec:model}
The question now is how to explain the observed collection patterns of discoverers. A possible explanation lies in the discoverers being more influential than the other users which in turn leads to the items collected by them eventually becoming popular. However, most of the systems that we analyze here lack any explicit mechanism for users to exert influence over the others, especially on such short time scales as we speak of here (we use $N_D=5$ through the paper, which means that only the first five users are awarded a discovery for collecting a relevant item). We have also data from Yelp.com, a web site for crowd-sourced reviews of local businesses, which is particular for comprising both bipartite user-item data and an explicit social network of users. However, we find no correlation between the number of friends and user surprisal which indicates that even when explicit influence can be exerted, it is not sufficient to explain the behavior of discoverers (see SI, Section S3, for details). This agrees with the finding that easily influenced individuals contribute to the rise of exceptionally popular items more than so-called influentials~\cite{watts2007influentials}. In the Amazon data, we also have the information on the number of users who find a review useful, which allows us to study the possible correlation between the average level of usefulness of a user's reviews and the user's surprisal value. However, we find no significant correlation which suggests that well-written and informative reviews do not contribute to the success of discoverers.

Motivated by these observations as well as by the presence of discoverers across many different systems, we propose an intrinsic mechanism to explain the observed discovery patterns. We first assume that some items are inherently more fit for a given system than the others and thus have higher chance of becoming very popular in the long run. Network models with node fitness have been studied in the past~\cite{bianconi2001competition,yook2002modeling,caldarelli2002scale} and they have been used to model various systems such as the World Wide Web~\cite{kong2008experience}, citations of scientific papers~\cite{Medo11,wang2013quantifying}, and an online scientific forum~\cite{medo2014statistical}, for example. Unlike the existing models, we then assume that the users differ in how they perceive item fitness and choose the items for their collections. While the first group of users are driven by item popularity and thus mostly ignore new and little popular items, the second group of users are driven by item fitness. Discoverers then emerge among the users in the latter group because: (1) fitness-driven users are consistently among the first ones to collect items of high fitness, (2) high fitness items often become very popular, (3) active fitness-sensitive users have the potential to achieve many discoveries and eventually be identified as discoverers by the statistical procedure that we propose here.

Two groups of consumers---innovators and imitators---are considered also by the Bass model~\cite{bass1969product} which constitutes a seminal model for the diffusion of innovations. However, the Bass model does not consider competition among the items and the link between an item's final popularity and its properties. Because of the focus on individual items, this model also does not consider the question whether individual users repeatedly act as innovators or imitators, respectively. Finally, the Bass model predicts a temporal exponential decay of the number of links received by a node which disagrees with the linking patterns of real systems where the temporal decay is slower (see \cite{Wang13} for a quantitative study of the Bass model in scientific citation data and a comparison with the network growth model proposed in \cite{Medo11}). We thus do not attempt to use the Bass model for modeling the discovery patterns found in real data.

We generate artificial bipartite networks with $U$ users where the number of items gradually grows from a small number $I_0$ to $I$ (we use $U=4000$, $I_0=50$, and $I=8000$ here). There are $U_F$ fitness-sensitive users and the remaining $U-U_F$ users are popularity-driven. Each user is further endowed with a level of activity which determines how likely is the user to collect a new item in any time step. While one can vary the distribution of activity among the users to model a broad range of real systems, user activity values are for simplicity drawn from the uniform distribution $[0,1]$ here. Item fitness quantifies how suitable and attractive is an item to the given system and its users; fitness values $f_{\alpha}$ are drawn from the power-law distribution with the lower bound $f_{\mathrm{min}} = 1$ and exponent $3$. As the subsequent analytical computation shows, a power-law fitness distribution directly translates into a power-law distribution of item popularity. Our choice of the item fitness distribution thus allows us to mimic real systems where the distribution of node popularity (degree) is often broad, typically power-law or log-normal~\cite{clauset2009power}. Time at which item $\alpha$ has been added in the system is denoted as $\tau_{\alpha}$. New links are added regularly until the final network density $\eta$ is achieved; the total number of links is thus $L=\eta U I$. To reach $I$ items before all links have been added in the network, new items are added every $L / (I - I_0 + 1)$ steps.

In the simulation, one user-item link is added in every time step. The user who creates this link is chosen from the pool of users with probability proportional to user activity. If a fitness-driven user $i$ creates a link at time $t$, the probability of choosing item $\alpha$ is proportional to
\begin{equation}
P_{i\alpha}\sim f_{\alpha}\,A(t - \tau_{\alpha})
\end{equation}
where $A(t - \tau_{\alpha}) = \exp[-(t - \tau_{\alpha}) / \theta]$ is an aging factor (see \cite{Medo11,medo2014statistical} for the original model of network growth with heterogeneous fitness and aging). Consequently, $\theta$ is a typical lifetime at which item attractiveness decays; we use $\theta=1000$ which is neither too quick (in which case the high-fitness items do not have sufficient time to attract many links and the resulting degree distribution is thus very homogeneous) nor too slow (in which case a strong bias towards old items develops and the fitness-popularity correlation is low). If a popularity-driven user $i$ creates a link at time $t$, the probability of choosing item $\alpha$ is proportional to
\begin{equation}
P_{i\alpha}\sim (k_{\alpha}(t) + 1)\,A(t - \tau_{\alpha})
\end{equation}
where $k_{\alpha}(t)$ is the degree (popularity) of item $\alpha$ at time $t$. The additive term in $k_{\alpha}+1$ is necessary to allow items of zero degree (every item is introduced in the system with zero degree) to gain their first links. Multiple links between a given user and an item are not allowed.

\subsection{Basic analytical results}
Denoting the fraction of fitness-driven users as $\mu_F := U_F/U$, we can write the following continuum equation which describes the evolution of the average degree of item $\alpha$ (see~\cite{albert2002statistical,Medo11} for more details on the continuum approximation approach)
\begin{equation}
\label{master2}
\begin{aligned}
\frac{\partial\avg{k_{\alpha}(t)}}{\partial t} =\,\, & 
\mu_F\,\frac{f_{\alpha}\,A(t-\tau_{\alpha})}{\sum_{\beta}f_{\beta} A(t-\tau_{\beta})}\, +\\
& +(1-\mu_F)\,\frac{(k_{\alpha}(t)+C)A(t-\tau_{\alpha})}{\sum_{\beta}(k_{\beta}(t)+C) A(t-\tau_{\beta})}
\end{aligned}
\end{equation}
where the two terms represent the contribution of the fitness- and popularity-driven users, respectively. The presence of the aging factor $A(\cdot)$ allows us to replace the sums in fraction denominators with their average values to which the sums approach at the time scale given by the form of $A(\cdot)$ and then fluctuate around them. In particular, we have
\begin{equation}
\label{mf}
\begin{aligned}
\sum_{\beta}(k_{\beta}(t)+C) A(t-\tau_{\beta})\to& \Omega_k,\\
\sum_{\beta}f_{\beta} A(t-\tau_{\beta})\to& \Omega_f.
\end{aligned}
\end{equation}
Equation (\ref{master2}) can now be solved analytically and yields the asymptotic result
\begin{equation}
\avg{k_{\alpha}(\infty)} = \biggl(\mu_F\,\frac{f_{\alpha}}{\Omega_f}+(1-\mu_F)\frac{C}{\Omega_k}\biggr)\,
\frac{\mathrm{e}^{(1-\mu_F)T/\Omega_k}-1}{(1-\mu_F)/\Omega_k}
\label{final}
\end{equation}
where $T=\int_{0}^{\infty}A(t)\,\mathrm{d} t$. Results for the previous model with preferential attachment, fitness and aging presented in~\cite{Medo11,Wang13} are recovered by setting $\mu_F=0$ and replacing $T$ with $Tf_{\alpha}$. We see that the expected final degree of items is indeed proportional to item fitness (the proportionality factor is given by the fraction of leaders in the system).

\begin{figure*}
\centering
\includegraphics[scale = 0.66]{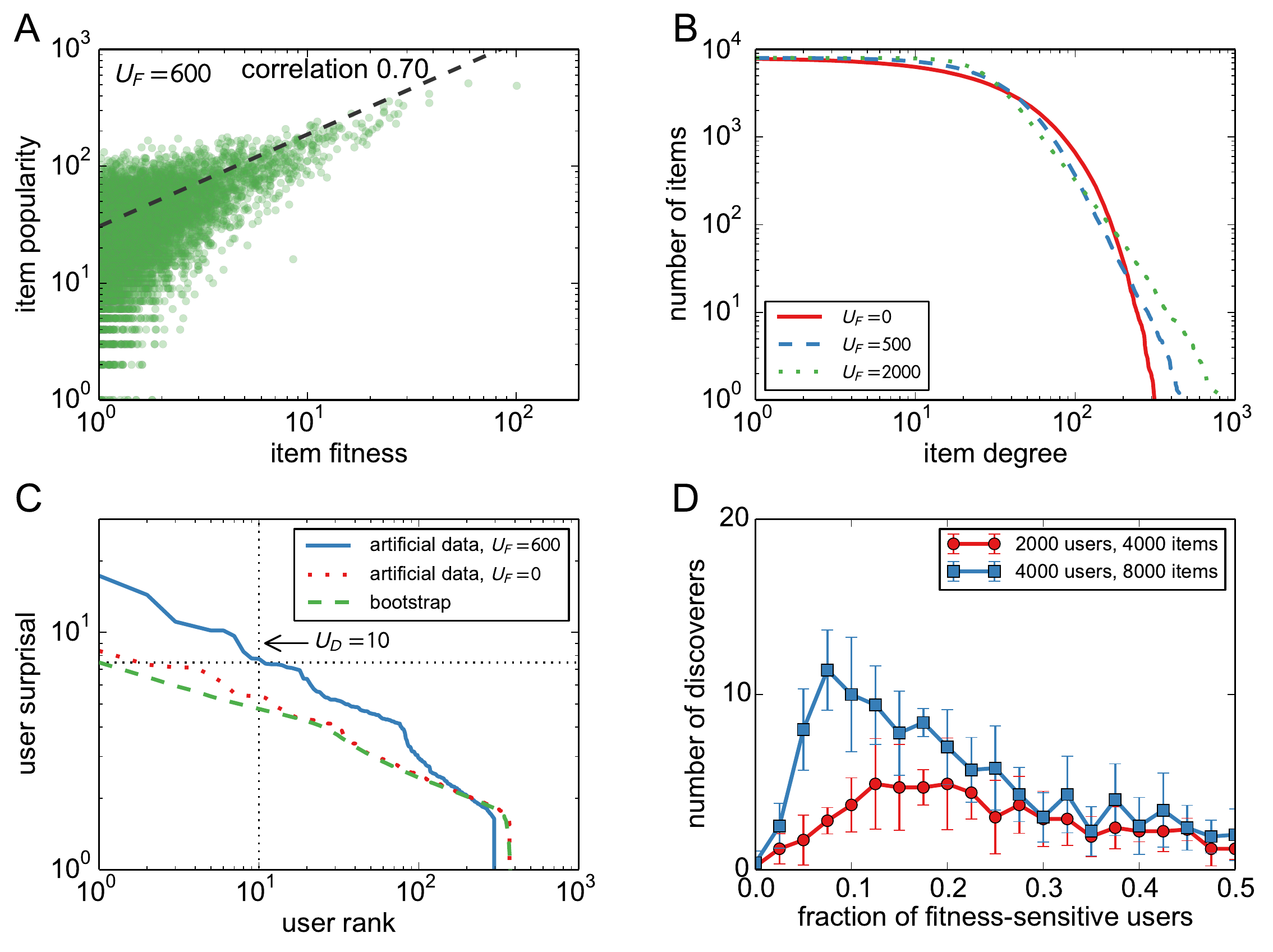}
\caption{User surprisal and discoverers in artificial model networks. (A) The relation between item fitness and popularity as well as the cumulative distribution of item popularity ($U_F=600$). (B) Item degree distributions for various values of $U_F$. (C) The Zipf plot of user surprisal in artificial data ($U_F=0,\ 600$) and in bootstrap (this curve is independent of $U_F$). (D) The dependence of the number of discoverers identified by the proposed statistical procedure on $U_F/U$. For comparison, we also show here results for 2,000 users and 4,000 items.}
\end{figure*}

We finally note that one can devise a model with continuously-distributed user ability $a_i\in[0,1]$ where the two aforementioned item-choosing equations can be merged in one. We have studied the multiplicative form
\begin{equation}
P_{i\alpha}\sim f_{\alpha}^{a_i}\,(k_{\alpha}(t) + 1)^{1-a_i}\,A(t - \tau_{\alpha})
\end{equation}
which implies that users of ability one respond only to item fitness, users of ability zero respond only to item popularity, and there is a continuous spectrum of user behavior between these two boundary ability values. However, we find the binary model with two discrete user groups easier to interpret and more amenable to analytical solution.

\subsection{Results on model data}
Simulation results for the artificial model are presented in Figure 4. Figure 4A shows that when a significant number of users are sensitive to item fitness (here $U_F=600$), the resulting networks exhibit strong correlation between item fitness and popularity. As $U_F$ decreases, this correlation gradually vanishes because we assume that the popularity-sensitive users ignore item fitness. As shown in Figure 4B, the distribution of item popularity is indeed rather broad and displays a power-law tail when $U_F$ is positive which agrees with the approximate analytical solution above. Figure 4C demonstrates that when $U_F$ is positive, user surprisal computed in model data differs from the bootstrap surprisal profile in the same way as we have shown in Figure 2 for the real data. The number of identified discoverers as a function of the number of fitness-sensitive users is displayed in Figure 4D. The dependence is notably non-monotonous. When $U_F$ is small, the correlation between item fitness and popularity is low and many of the popular items that are used to assign discoveries are thus of low fitness; the fitness-sensitive users thus fail to achieve many discoveries and the resulting $U_D$ is close to zero. As $U_F$ grows, the fitness-popularity increases and so does $U_D$ but eventually, there number of fitness-sensitive users is too large for the number of available discoveries and $U_D$ declines. For intermediate values of $U_F$, the numbers of identified discoverers are significant and we can thus conclude that the proposed simple model is able to reproduce the discovery patterns observed in real data. The observed fraction $U_D/U$ which gets as high as $0.03\%$ at $U_F=300$ is similar to that found in the Amazon data.

Note that the groups of fitness-driven users and discoverers are in general not the same. While in the current setting, all discoverers identified using the proposed statistical framework are fitness-driven, only a small fraction of fitness-driven users are identified as discoverers (in Figure 4D, for example, $U_F=600$, yet $U_D\lesssim 10$). There are various reasons why a fitness-driven user does not become a discoverer: the user is not active enough, or by chance becomes active at moments when there are no relevant items (that is, little popular high-fitness items) available and hence no discoveries can be made, or simply fails to connect with the available relevant items because of the probabilistic network growth mechanism. The fact that discoverers are found in the model data is thus not automatic and the number of statistically significant discoverers depends strongly on model parameters.

\begin{figure*}
\centering
\includegraphics[scale = 0.66]{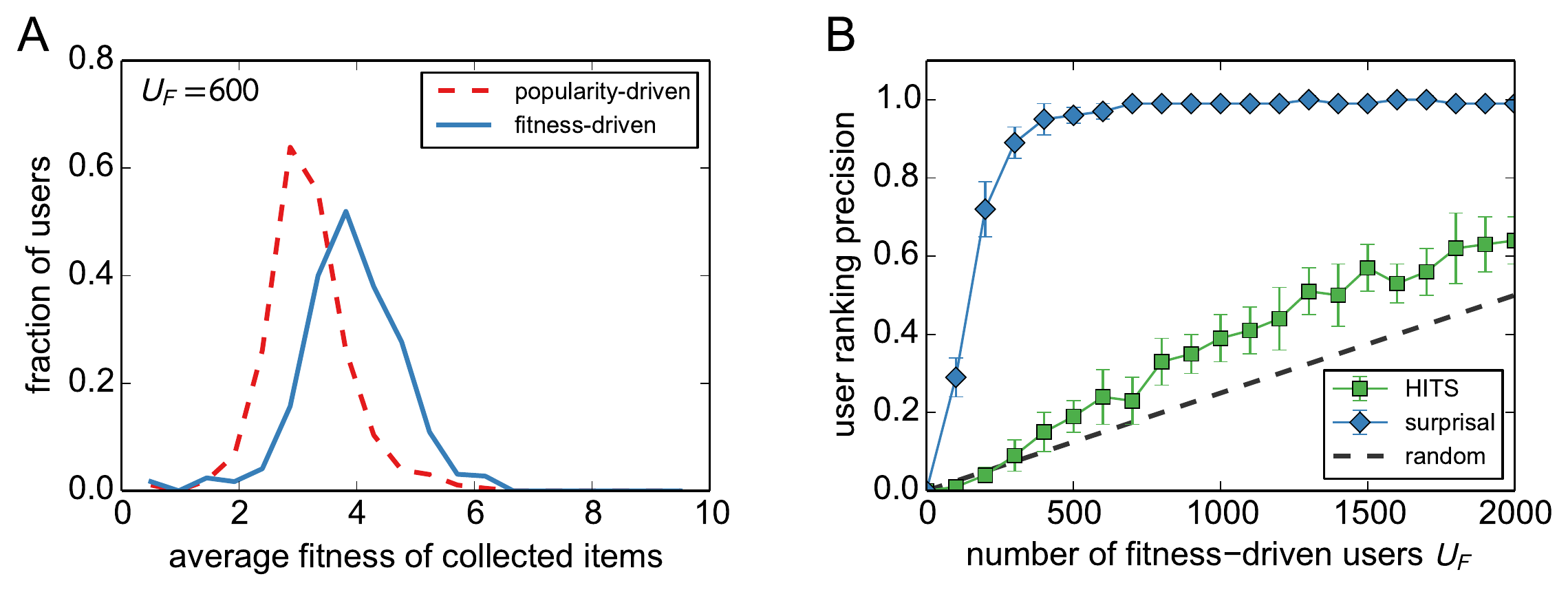}
\caption{Ranking of users in model networks. We show here results for networks with 4,000 users and 8,000 items. (A) Histograms of the average fitness of items collected by fitness- and popularity-driven users, respectively, overlap substantially ($U_F=600$, one network realization). (B) The performance of HITS and surprisal in ranking users according to their sensitivity to item fitness. Ranking precision is defined as the fraction of fitness-driven users in top 50 positions of the ranking (results are averaged over 10 model realizations, error bars represent the standard deviation).}
\label{fig:algorithms}
\end{figure*}

We close with a discussion of profound implications of the presence of fitness- and popularity-driven users on node ranking algorithms in networks. The typical goal of a network ranking algorithm is to find the most important nodes. In the context of a user-item network, the most important item nodes are those that have the highest fitness and the most important user nodes are those who are fitness-driven. To go beyond nodes' local neighborhoods and thus benefit from the network structure, these algorithms usually allow scores to propagate between nodes~\cite{langville2005survey,franceschet2011pagerank}. In the context of bipartite user-item networks, this means that the user score is given by the score of the items that have been collected by the user and the item score score is given by the score of the users who have collected the item, as formalized by a bipartite version~\cite{deng2009generalized} of the classical HITS algorithm~\cite{kleinberg1999authoritative}. The present class of model data however poses an important difficulty: Figure 5A shows that despite some users being more sensitive to item fitness than the others, both user groups ultimately collect items of similar fitness (in other words, the correlation between user ability and the average fitness of collected items is low). The problem is particularly pronounced for the users who have collected the best items on average: only $30\%$ of the 50 users who are best in this respect are actually fitness-sensitive; the remaining $70\%$ are popularity driven. The reason for this is simple: after fitness-driven users find high fitness items, popularity-driven users driven by popularity are likely to copy their choice and end up with items which are only marginally worse than those collected by the fitness-driven users.

This observation suggests that the broad class of network-based algorithms may be unsuitable for networks where preferential attachment allows ordinary users to effectively copy the choices made by discoverers. To verify that, we apply a bipartite HITS algorithm on artificial networks and evaluate the fraction of top 50 positions in thus-produced user ranking which is actually occupied by fitness-sensitive users. The algorithm assigns scores $x_i$ and $y_{\alpha}$ to users and items, respectively, that satisfy the set of equations
\begin{equation}
\label{HITS}
x_i = \frac1{k_i}\sum_{\alpha\in\mathcal{I}_i} y_{\alpha},\quad
y_{\alpha} = \sum_{i\in\mathcal{U}_{\alpha}} x_i
\end{equation}
where $\mathcal{I}_i$ is the set of items collected by user $i$ and $\mathcal{U}_{\alpha}$ is the set of users who have collected item $\alpha$. In other words, the score of users is given by the average score of items collected by them and the score of items is given by the total score of users who collect them. The set of equations is typically solved iteratively by first setting uniform score vectors and then recomputing $x$ values based on the current $y$ values and vice versa~\cite{medo2013network}. To prevent the score vectors from diverging, they need to be normalized after each iteration.
Figure 5B confirms that bipartite HITS indeed performs only marginally better than random ranking of users (the fraction of fitness-driven users in top 50 is then the same as in the whole population, i.e. $U_F/U$). We now see that the failure of this algorithm is in its ignorance of the time information---users who discover genuinely valuable content are thus indistinguishable from those who later copy their choice. Omitting, for example, the normalization with $k_i$ in Eq.~(\ref{HITS}) thus does not change the results significantly. To correctly rank users in this network, an algorithm needs to account not only for who has collected what but also when they have done so. Although surprisal has not been devised to rank users, we apply it in the artificial networks. Figure~\ref{fig:algorithms}B shows that the precision of the rankings obtained by user surprisal exceeds that achieved by bipartite HITS and reaches near-perfect precision of 95\% at $U_F = 400$.

It has been demonstrated that in real systems, the popularity of items is path-dependent and sensitive to system design and possible external factors~\cite{salganik2006experimental,muchnik2013social}, which questions the basic premise of the proposed user surprisal measure which uses the most popular items as the relevant items for which discoveries are awarded to users. The analysis of model data allows us to return to this important point equipped with better understanding of both the statistical procedure and the systems on which it is applied. We find discoverers in the model data despite the fact that the correlation between item fitness and popularity is far from perfect (see Figure 4A) and almost all of the identified discoverers are indeed fitness-sensitive (see Figure~\ref{fig:algorithms}B for $U_F\gtrsim 400$). This high robustness towards sub-optimal choice of relevant items is due to the fact that when some popular items are actually of low-fitness, fitness-sensitive users simply ignore them. By contrast, the popularity-sensitive users gain some discoveries for these inferior popular items but since these users are typically in majority by a wide margin, thus-achieved discoveries are not sufficient to achieve significant values of user surprisal. We see that while the imperfect choice of the relevant items thus reduces the signal for fitness-sensitive, it creates only a weak false signal for popularity-sensitive users.

\section{Discussion}
In this article, we introduce discoverers as the users in data from real systems who significantly outperform the others in the rate of making discoveries, \emph{i.e.} in being among the first ones to collect items that eventually become very popular. We develop a statistical framework to identify the discoverers and use it to demonstrate that they can be found across a number of online systems where users have the freedom to choose from a large number of possible items. The proposed approach is suitable to any data with time information. Evidence for discovery behavior in monopartite networks (work in progress) shows that our approach is applicable and relevant to an even broader range of systems than those studied here. The ability to identify the discoverers is shown beneficial for predicting the future popularity of items as well as for ranking the users. Motivated by the generality of the observed phenomenon and a lack of direct ways for an individual to influence other users in the systems studied here, we search for a unifying mechanism to model the discovery behavior. To this end, we generalize the preferential-attachment network growth model with fitness and aging~\cite{Medo11} by assuming that not only the item nodes differ in their fitness but also the user nodes differ in their sensitivity to item fitness. In the model data, fitness-sensitive users recognize the high fitness items, collect them, and these items then often eventually become very popular due to their high fitness. While the model reproduces the discovery patterns found in the real data, we emphasize that the main goal of the model is to show that the reported discovery patterns can be modeled based on a small variation of the existing network growth models. A comprehensive study of model parameterizations that best agree with real data as well as both quantitative and qualitative analysis of various possible reasons for the presence of discoverers in real data remain as future research challenges.

Model data show low correlation between user ability and the average fitness of items collected by them. This seemingly ordinary finding has far-reaching implications because it contradicts the basic assumptions of many network-based ranking algorithms such as PageRank and HITS. We show that while a traditional ranking algorithm indeed performs poorly on the model data, the newly developed user surprisal metric works well due to the fact that it takes the system's complete time evolution into account, not just the final state. This indicates the need for new algorithms that act on detailed network representations with full information about the creation time of all nodes and links. A similar observation in the context of diffusion in temporal networks~\cite{scholtes2014causality} indicates that the suggested approach is a general one: evolving networks require temporal methods.

We stress again that the classical concepts of social leaders or innovators who have high social status or are well positioned in the social network, extensively studied in the past~\cite{katz1955personal,rogers2010diffusion,bass1969product,aral2012identifying}, do not provide a full explanation for the presence of discoverers who do not share any advantageous or privileged position and achieve discoveries consistently over time. Our work demonstrates the presence of discoverers in social systems and at the same time calls for a deeper understanding of their behavior and roles. To quantify the level to which a user's discovery performance is due to some external influence (for example, a minority of the identified high surprisal users in the Amazon data are members of the Amazon's Vine Voice program which gives them advance access to not-yet-released products) is just one of the steps towards understanding the phenomenon of discoverers.

Our results provoke several further questions. The null hypothesis of the discovery probability independent of time assumes the real discovery rate is constant. While Figures S3 and S4 show that this requirement is fulfilled in the studied datasets, a general framework where the system's timespan is divided into multiple bins and $p_D$ is computed for each bin separately could render the requirement unnecessary. We have already discussed that the proposed statistical framework is robust to sub-optimal choice of the set of relevant items, which is crucial for its relevance in any realistic setting. However, we may still attempt to improve the initial choice on the basis of the computed values of user surprisal. For example, we can replace the popular items that are actually neglected by the users of high surprisal with the items that are better received by them. User surprisal can be then recomputed using the updated group of relevant items, and by repeating the described steps eventually obtain a closed self-consistent system where relevant items are those that are collected by high surprisal users and high surprisal users are those who early discover the relevant items.  Regarding the proposed network growth model, an extensive set of dynamical and statistical measurements is needed to determine how well the model represents the relevant features of real systems in comparison with other existing network models. For example, if we want to produce realistic discovery patterns, is it necessary to assume that the popularity-driven users are wholly ignorant of item fitness? At a more basic level, the knowledge of user surprisal values gives us for the first time the possibility to discriminate between similarly popular items purely based on their success among users of different surprisal. This and the potential use of discoverers for predicting the future success of items illustrated in Figure 3 are the first hints of our work's potential applications in e-commerce and marketing in general.

\begin{acknowledgments}
This work was supported by the Swiss National Science Foundation Grant No. 200020-143272 and by the EU FET-Open Grant No. 611272 (project Growthcom). We thank Stanislao Gualdi for early discussions, Zike Zhang for providing us with the e-commerce keyword data, and Alexandre Vidmer for providing us with the Yelp data.
\end{acknowledgments}

\bibliographystyle{apsrev4-1}
\bibliography{leaders}

\onecolumngrid
\appendix

\newpage
\begin{center}
\bfseries\large Supplementary Information
\end{center}

\section{Data description}
We analyzed six different real data sets. The description and results for the first two, Amazon and Delicious, are presented in the main text. Here follows a brief description of the other four data sets. The corresponding results are presented in Figure S2.

\begin{enumerate}
\item We downloaded the \url{Epinions.com} consumer review data from \url{konect.uni-koblenz.de/networks/}. The original data comprise 120,492 users, 755,760 items and 13,668,320 ratings. Time span of the data is from 9 January 2001 to 29 May 2002. In the raw data, the time stamps exhibit a periodic pattern with respect to link order. In addition, many links appear at the starting day of the data. To avoid these two problems, we use only links ranked from 12,276,827 to 13,213,749 in the original data. Since ratings are in the integer scale from 1 to 5, we apply the same threshold mechanism as in the Amazon data. The final subset contains 17,542 users, 32,482 items and 753,392 links. Time span of the subset is from 16 January 2001 to 29 May 2002 (499 days in total).

\item Keyword data from the biggest Chinese online shopping website \url{taobao.com} were crawled via open API from the web site. In the Taobao e-commerce platform, vendors can use keywords to describe their products and well-chosen keywords can contribute to their products being ranked at the top of customers' search results. At the same time, vendors have to pay a price for using keywords and the price of a keyword depends on the keyword's popularity---vendors thus have an incentive to invent new keywords or early adopt already existing keywords. The data comprise 2,824,853 links between 1,523 online retailers and 915,271 keywords that they attached to their products. Time span of the data is from 12 November 2009 to 21 June 2014 (40,360 hours in total).

\item Movielens movie rating data were obtained from \url{grouplens.org/datasets/movielens/}. The original data comprise $10,000,054$ ratings from $71,567$ users to $10,681$ movies in the online movie recommender service MovieLens. Since ratings are in the integer scale from 1 to 5, we apply the same threshold mechanism as in the Amazon data. Time span of the data is from January 1995 to January 2009 (122,634 hours in total). We use the subset from hour 40,000 until the end of the data to avoid an initial period of low user activity. The final subset contains $2,132,128$ links between $44,548$ users and $7,974$ items.

\item Netflix DVD rating data were made available for the Netflixprize contest and can be still downloaded from \url{www.netflixprize.com/}. The original data comprise $100,481,826$ ratings from $480,189$ users to $17,770$ movies in the online DVD rental website Netflix. Since ratings are in the integer scale from 1 to 5, we apply the same threshold mechanism as in the Amazon data. Time span of the data is from January 2000 to January 2006 (2242 days in total). We use the subset from day 500 to day 1,500 to constraint the data size. The final subset contains $2,775,772$ links between $115,131$ users and $7,351$ items.
\end{enumerate}

When taking a subset, we only include objects that have not appeared before the subset's beginning, such that we do not misrepresent some later links for discoveries (if item X first appear at time 100 and collects many links until a subset's beginning, we do not want to award discoveries to the first links to item X within the subset because they are actually no discoveries). Creating a subset still means that we include only a part of each user's links. As a result, a user could be a good discoverer in our subset but a less good one before or after. However, this is unbiased sampling from a user's links which does not give undue advantage to anyone.

\section{Evaluation of future degree evolution}
We choose subsets of time span $T_S$ by choosing their starting time $T_X$ at random from the range $[0, T_W - T_S - T_F)$ where $T_W$ is the time span of the whole dataset and $T_F$ is the length of the future time window, over which we observe the future degree increase of items (see the diagram below). Each subset contains only items that have not been collected before the subset's starting point (if we would include those items, we could assign false discoveries for them despite the fact that those items might have already collected a substantial number of links before the subset's start). A given subset is then used to compute surprisal of all its users. We further choose all items that have received exactly one link and they have appeared at most $\tau_{\mathrm{max}}$ before the subset's end time (this represents young and yet unpopular items) as items of interest. $\tau_{\mathrm{max}}$ is $20$ and $2$ days for the Amazon and Delicious data, respectively, which accounts for different dynamics of these two systems.

We then track all links that are attached to the items of interest in the future time window of length $T_F$ (i.e., these links are not part of the subset which was used to compute user surprisal values). This allows us to compute the average degree of these items as a function of time. Results are further averaged over 100 subsets defined by their $T_X$ value. In Figure 2 in the main text, we plot the average degree increase computed separately for items of interest collected by three distinct user groups: users of zero surprisal, users of low (less than ten) surprisal and users of high (ten and more) surprisal. The chosen parameters of this evaluation procedure are summarized in Table S3. While their values influence the detailed shape and relative height of the curves reported in Figure 2, the main result of items collected by high surprisal users being more popular than items collected by users of zero or low surprisal holds always.

\vspace*{12pt}
\begin{center}
\includegraphics[scale = 0.9]{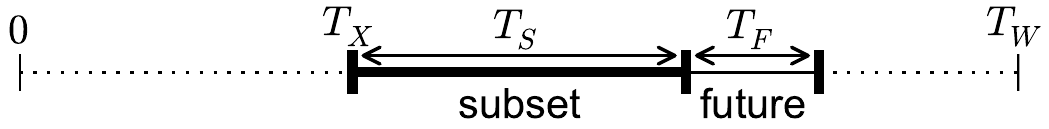}\\
A diagram of the subset creation for the evaluation of future degree evolution.
\end{center}
\vspace*{12pt}

\section{Analysis of the Yelp data}
We now address the question of whether the observed discovery patterns can be explained by the social influence of users. To this end, we use a dataset from the  Yelp academic challenge, round 4 (see \url{http://www.yelp.com/academic-dataset} for more information). The advantage of this dataset is that it features both the bipartite user-item network as well as the social user-user network (the Delicious web site also allowed the users to form friendship links but unfortunately we do not have the social network information and thus have to use a new dataset). The input data contains  252,898 users, 42,153 items (which in this case represent businesses), 955,999 friendship links, and 1,125,458 reviews in the integer scale from 1 to 5; the time stamps run from 0 to 3558 (measured in days). We only keep the users who have at least one friend and one authored at least one review. As for the other datasets, we use the rating threshold of four and focus on a subset of the data (in this case the the evaluations from days 1000 until 3499; we thus ignore the rather long initial period of 1000 days which aims at avoiding the notorious items that existed before day 0 and awarding discoveries for them would therefore be unjust). We finally have a dataset with 80,840 users, 33,661 items, 348,060 user-item links and 674,231 directed user-user links.

As in the other reported datasets, also the Yelp data features discoverers: the largest user surprisal value is $21$, the average highest surprisal in bootstrap realizations is $9.7$, and the number of identified discoverers is $30$. A comparison of the set of 100 highest surprisal users with the set of 100 most social (as measured by the number of friends) users reveals that the two sets share only one user and even this user actually does not pass the bootstrap surprisal threshold (the user's surprisal value is thus not statistically significant and could happen by chance). We can conclude that in the Yelp data, users with many social contacts are in now way more successful in achieving discoveries than users with few social contacts.

\newpage

\section*{Supplementary Tables}

\begin{center}
{\small
\begin{tabular}{rrrrrrrrrr}
\hline
  Dataset &   Users &     Items &     Links &    Time span & $\overline{k_i}$ & $K_i$ & $\overline{k_{\alpha}}$ & $K_{\alpha}$ & $k_D$\\
\hline
   Amazon & 406,275 &    76,205 &   713,581 &   3,000 days &  1.8 &  1,296 &  9.4 &    790 & 127\\
Delicious & 107,810 & 2,435,912 & 9,322,949 & 20,000 hours & 86.5 &  6,582 &  3.8 &  7,014 & 40\\
 Epinions &  17,542 &    32,482 &   753,392 &     499 days & 42.9 &  5,809 & 23.2 &    508 & 154\\
  Keyword &   1,523 &   915,271 & 2,824,853 & 40,360 hours & 1855 & 23,775 &  3.1 &    227 & 32\\
Movielens &  44,548 &     7,974 & 2,132,128 & 82,624 hours & 47.9 &  2,419 &  267 & 18,858 & 3,852\\
  Netflix & 115,131 &     7,351 & 2,775,772 &   1,000 days & 24.1 &    959 &  378 & 26,700 & 8,256\\
\hline
\end{tabular}}
\end{center}
Table S1: \textbf{Basic statistical properties of the studied datasets.} The time span column specifies both duration and time resolution of the datasets. $\overline{k_i}$ and $\overline{k_{\alpha}}$ are the mean user and item degree, respectively. $K_i$ and $K_{\alpha}$ are the largest user and item degree, respectively. $k_D$ is the smallest degree upon which an items is considered as one of items that are to be discovered when $f_D = 1\%$.

\vfill

\begin{center}
{\small
\begin{tabular}{rrrrrr}
\multicolumn{6}{c}{\normalsize Amazon}\\[2pt]
\hline
Rank & $k_i$ & $d_i$ & $r_i$ & $P_i^0$ & $s_i$\\
\hline
1  & 188 & 59 & 51.6 & $10^{-82}$ & 187.6\\
2  & 244 & 50 & 33.7 & $10^{-59}$ & 135.3\\
3  & 217 & 35 & 26.5 & $10^{-38}$ & 86.4\\
4  & 237 & 26 & 18.0 & $10^{-24}$ & 54.4\\
5  & 190 & 24 & 20.8 & $10^{-24}$ & 53.8\\
6  & 364 & 26 & 11.7 & $10^{-19}$ & 43.5\\
7  & 185 & 18 & 16.0 & $10^{-16}$ & 36.1\\
8  &  73 & 11 & 24.8 & $10^{-12}$ & 27.6\\
9  &  41 &  9 & 36.1 & $10^{-12}$ & 26.4\\
10 &  60 & 10 & 27.4 & $10^{-12}$ & 26.2\\
11 &  12 &  6 & 82.2 & $10^{-11}$ & 23.8\\
12 &  42 &  8 & 31.3 & $10^{-10}$ & 22.4\\
13 & 432 & 18 &  6.8 & $10^{-10}$ & 21.7\\
14 &  47 &  8 & 28.0 & $10^{-10}$ & 21.5\\
15 &  99 & 10 & 16.6 & $10^{-9}$ & 21.1\\
16 &  31 &  7 & 37.1 & $10^{-9}$ & 21.1\\
17 &  51 &  8 & 25.8 & $10^{-9}$ & 20.8\\
18 &  35 &  7 & 32.9 & $10^{-9}$ & 20.1\\
19 &  23 &  6 & 42.9 & $10^{-9}$ & 19.2\\
20 &  71 &  8 & 18.5 & $10^{-8}$ & 18.1\\
\hline
\end{tabular}
\qquad
\begin{tabular}{rrrrrr}
\multicolumn{6}{c}{\normalsize Delicious}\\[2pt]
\hline
Rank & $k_i$ & $d_i$ & $r_i$ & $P_i^0$ & $s_i$\\
\hline
 1 & 3556 & 195 &  3.9 & $10^{-56}$ & 127.9\\
 2 &  768 &  88 &  8.2 & $10^{-50}$ & 115.1\\
 3 & 3835 & 192 &  3.6 & $10^{-49}$ & 112.7\\
 4 & 2124 & 136 &  4.6 & $10^{-47}$ & 106.8\\
 5 & 2625 & 141 &  3.9 & $10^{-40}$ & 91.0\\
 6 &  894 &  82 &  6.6 & $10^{-40}$ & 90.7\\
 7 &  639 &  69 &  7.7 & $10^{-38}$ & 87.0\\
 8 & 1019 &  85 &  6.0 & $10^{-38}$ & 86.7\\
 9 & 4585 & 185 &  2.9 & $10^{-35}$ & 80.2\\
10 &  395 &  47 &  8.5 & $10^{-28}$ & 64.2\\
11 & 1060 &  73 &  4.9 & $10^{-27}$ & 62.8\\
12 & 1177 &  73 &  4.4 & $10^{-25}$ & 56.6\\
13 & 1116 &  71 &  4.6 & $10^{-25}$ & 56.5\\
14 &  808 &  59 &  5.2 & $10^{-24}$ & 54.1\\
15 & 2355 & 104 &  3.2 & $10^{-23}$ & 52.7\\
16 &  622 &  50 &  5.8 & $10^{-22}$ & 50.3\\
17 &  573 &  48 &  6.0 & $10^{-22}$ & 50.2\\
18 &  292 &  35 &  8.6 & $10^{-21}$ & 48.6\\
19 &   75 &  21 & 20.1 & $10^{-21}$ & 48.3\\
20 & 1580 &  78 &  3.5 & $10^{-20}$ & 46.4\\
\hline
\end{tabular}}
\end{center}
Table S2: \textbf{Twenty users with the highest surprisal in the Amazon and Delicious data.} $k_i$ is the degree of user $i$, $d_i$ is the number of discoveries by user $i$, $r_i:=d_i / (p_Dk_i)$ is the ratio between the actual number of discoveries $d_i$ and the number of discoveries expected under the null hypothesis $p_Dk_i$, $P_i^0$ is the probability that user $i$ makes at least $d_i$ discoveries under the null hypothesis, and finally $s_i:=-\ln P_i^0$ is the corresponding surprisal (cf. Equations (1) and (2) in the main text).

\vfill

\begin{center}
{\small
\begin{tabular}{rrrrr}
\hline
  Dataset &        $T_W$ &       $T_S$ &       $T_F$ & $\tau_{\mathrm{max}}$\\
\hline
   Amazon &   3,000 days &  2,000 days &    100 days & 20 days\\
Delicious & 20,000 hours & 7,200 hours & 2,400 hours & 480 hours\\
\hline
\end{tabular}}
\end{center}
Table S3: \textbf{Parameters of the future degree evaluation procedure.} $T_W$ is the time span of the whole dataset, $T_S$ is the time span of subsets, $T_F$ is the future time window in which the degree of items of interest is observed, and $\tau_{\mathrm{max}}$ is the maximal age of an item of interest in a given subset.

\newpage

\section*{Supplementary Figures}

\begin{center}
\includegraphics[scale = 0.55]{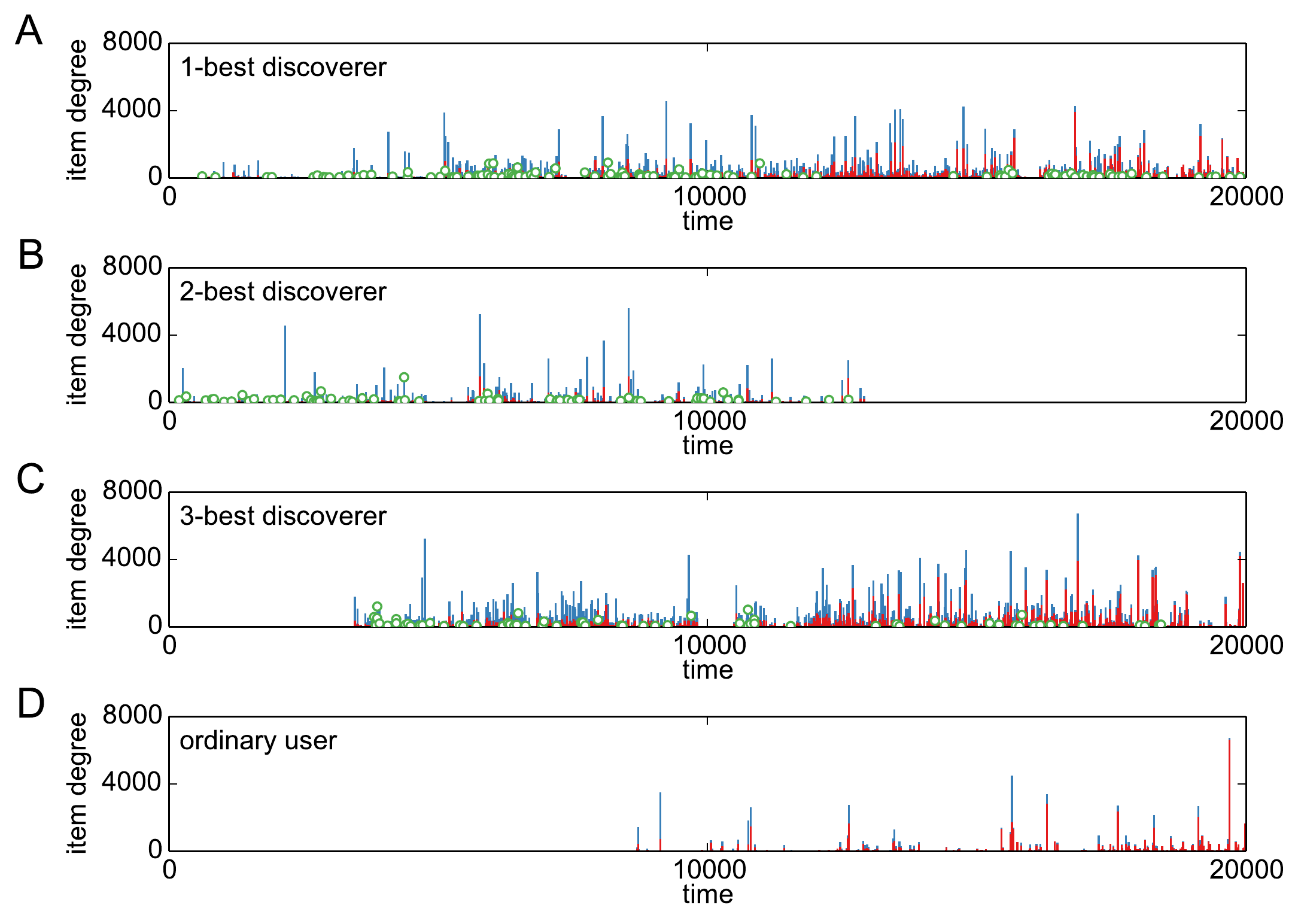}
\end{center}
Figure S1: \textbf{Collection patterns of users in the Delicious data.} Similarly as Figures 1A and 1B in the main text show the collection patterns of different users in the Amazon data, we do here the same for the Delicious data. Each bar again corresponds to an object collected at a given time point with the red and blue part indicating the item's degree at the moment when it was collected by a given user and the final degree, respectively. We show here data for three users with the highest surprisal and a randomly chosen active user without discoveries.

\begin{center}
\includegraphics[scale = 0.55]{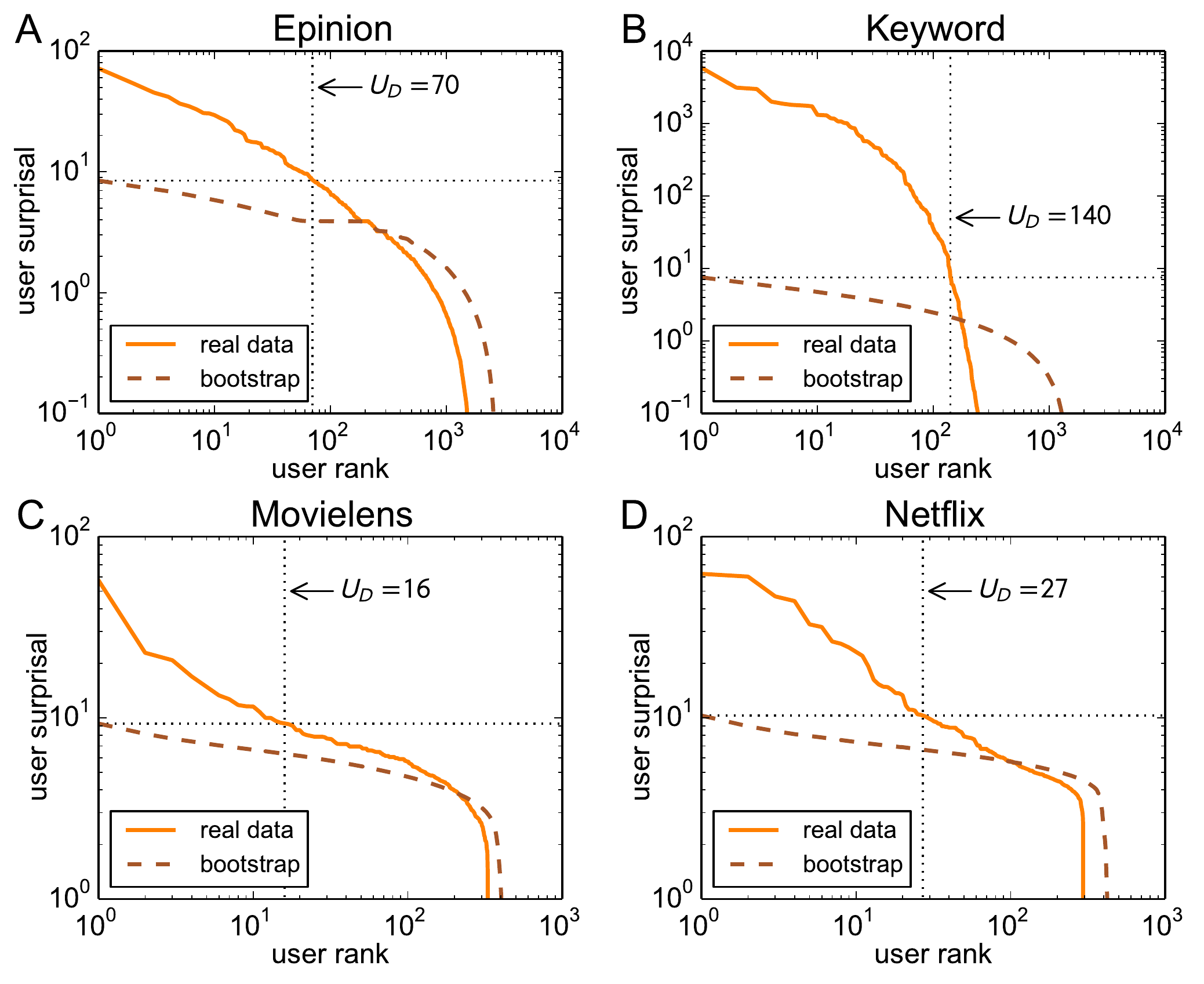}
\end{center}
Figure S2: \textbf{Bootstrap results for other data sets.} In analogy with Figure 1 in the main text, we show here Zipf plots of real and bootstrap surprisal values for four additional data sets (Epinions, Taobao keyword data, Movielens, and Netflix subsets). We do not indicate here the standard deviations of the bootstrap surprisal values because they are small---around $1$ for the top-ranked user and it decreases quickly with user rank.

\begin{center}
\includegraphics[scale = 0.55]{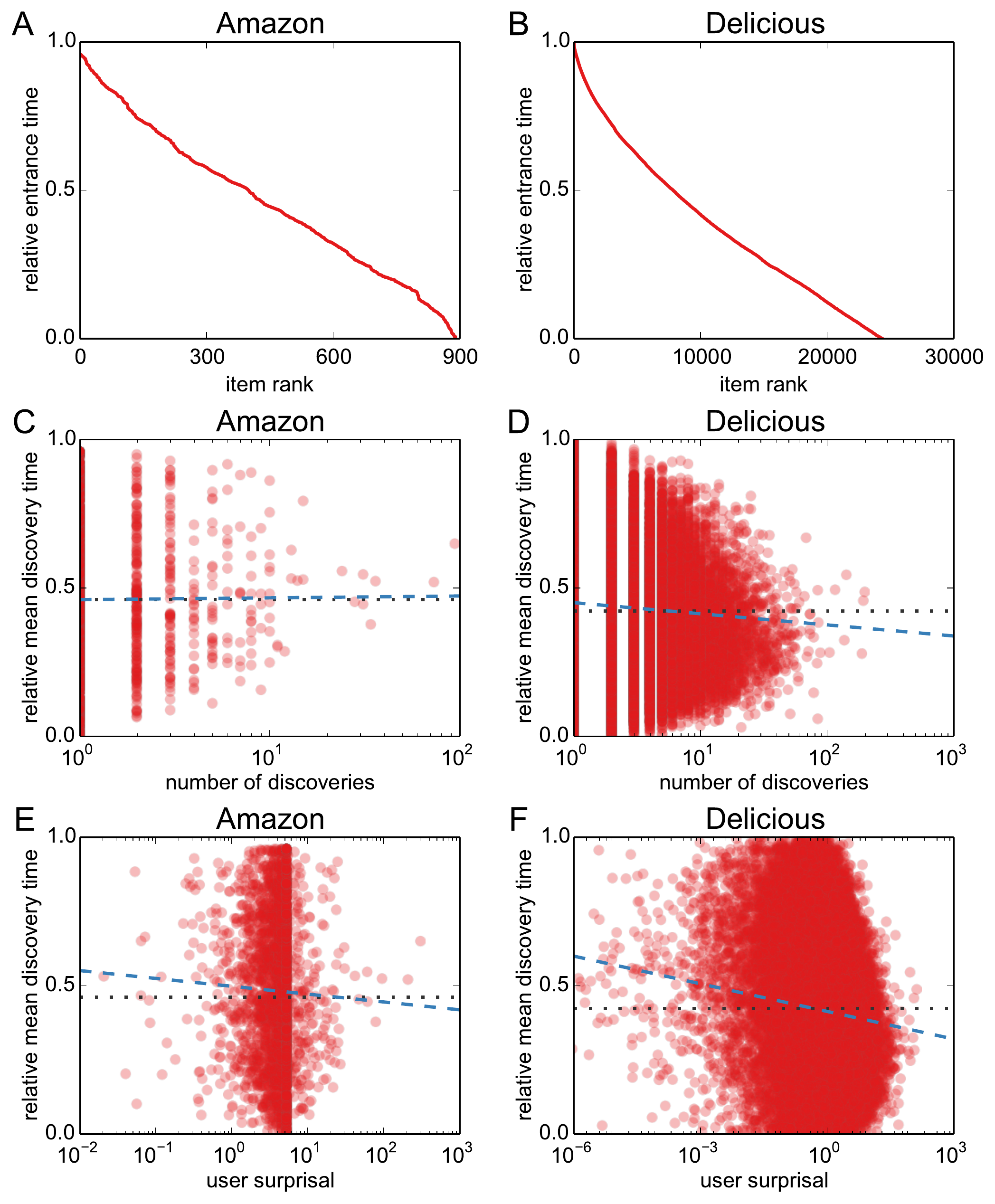}
\end{center}
Figure S3: \textbf{Temporal distribution of discoveries.} We investigate here whether discoveries and surprisal are not strongly biased towards, for example, the early period of the analyzed data when the number of items was small and thus it was maybe easier to make discoveries. To this end, we show the Zipf plots of the relative entrance time of target items (the relative entrance times of $0$ and $1$ correspond to the beginning and end of the data set). Nearly straight lines in panels A and B indicate that target items are distributed rather uniformly through the data time span. Panels C and D show the mean discovery time (again in relative units) of individual users plotted the number of discoveries achieved by them. The dotted line represents the mean discovery time averaged over all users. The dashed line represents a linear fit of the data for individual users in the log-linear plane. Panels E and F show the mean discovery time of individual users against their surprisal. See Figure S4 for more detailed information about the discovery patterns of top 20 users in both number of discoveries and surprisal.

\newpage

\begin{center}
\includegraphics[scale = 0.6]{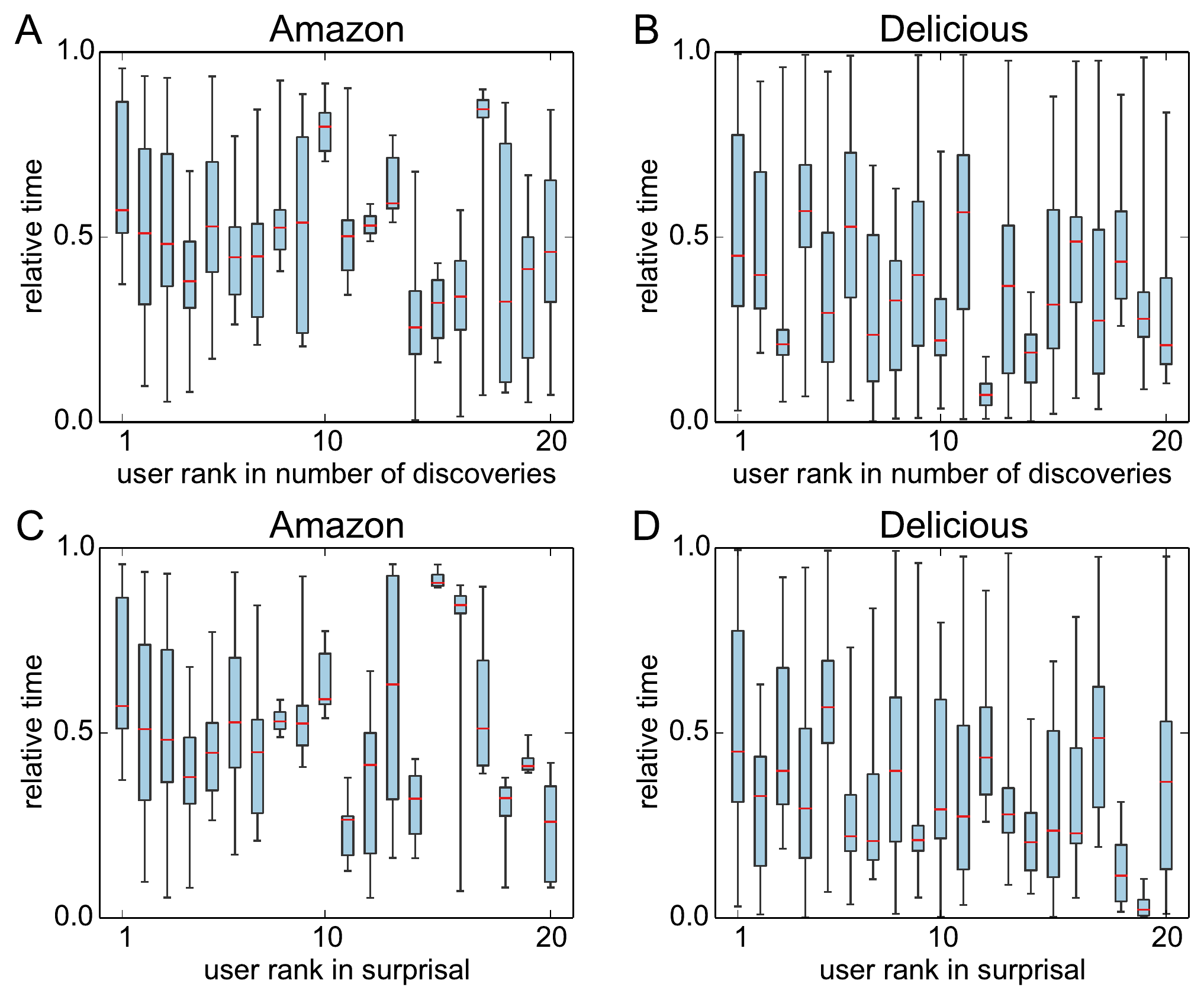}
\end{center}
Figure S4: \textbf{Temporal distribution of discoveries for top 20 users.} To uncover possible time bias in the discovery patterns of users, we show here box plots for discovery times by individual users who are ranked among the top 20 users either in number of discoveries (A, B) or in surprisal (C, D). The boxes represent the first and third quartile of the discovery times for each individual user; the bands represent median values; the whiskers represent the minimum and maximum values. One can see here that discoveries are spread over a substantial time period for majority of top users with only a few users achieving a substantial fraction of their discoveries at the very beginning of the data (the only exceptions are user \#19 in surprisal and user \#12 in number of discoveries, both in the Delicious data). We can conclude that high numbers of discoveries and surprisal achieved by some users are not due to their privileged position.

\newpage

\begin{center}
\includegraphics[scale = 0.6]{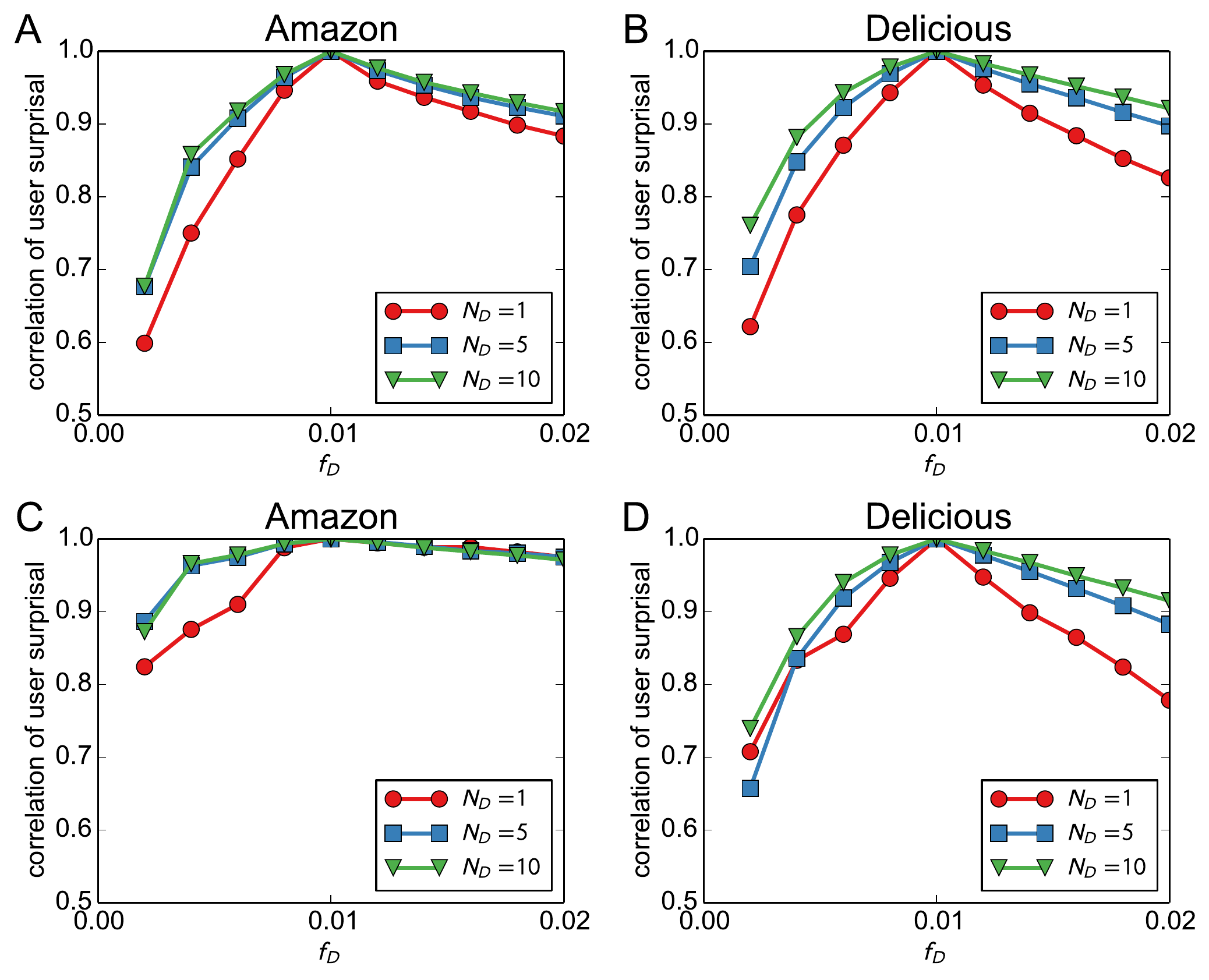}
\end{center}
Figure S5: \textbf{Stability of user surprisal values with respect to parameters.} In the main text, we choose $f_D^*=1\%$ and $N_D=5$ for both Amazon and Delicious data. To investigate the effect of these two parameters on user surprisal values, we first compute the vector of user surprisal values for $f_D^*$ and denote it $\vek{S}^*$. We then calculate the vector of surprisal values $\vek{S}$ for any different $f_D$ and compute the Pearson correlation coefficient $r(\vek{S}^*, \vek{S})$ which is then shown in panels A and B. The procedure is the same in panels C and D, except for the computation of Pearson correlation coefficient only over users whose surprisal in $\vek{S}^*$ is greater than $10$ (we focus in this way on users who matter most from the perspective of their discovery ability). Results for various values of $N_D$ (recall that $N_D$ first links attached to a target item are marked as discoveries) are shown here. When $N_D=1$, surprisal values are more sensitive to changes of $f_D$ because the information used to compute surprisal is then rather limited. Results (panels C and D in particular) show that surprisal values are rather robust: increasing or decreasing $f_D$ by the factor of two still yields correlation values above $0.9$ for both Amazon and Delicious data.

\newpage

\begin{center}
\includegraphics[scale = 0.6]{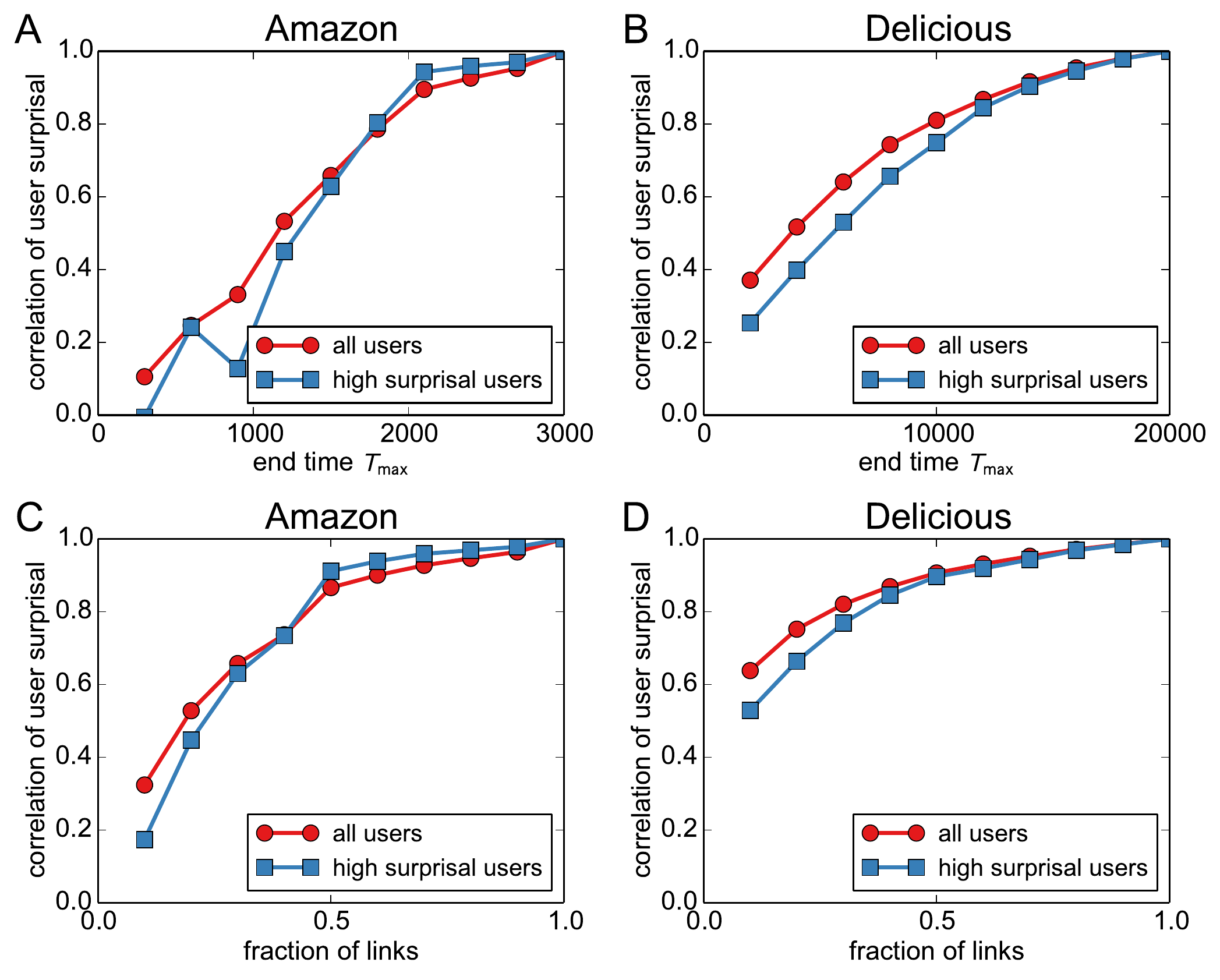}
\end{center}
Figure S6: \textbf{Stability of user surprisal values with respect to the data.} We compute here the vector of user surprisal values $\vek{S}^*$ using the full data and measure its correlation with the vector of user surprisal values $\vek{S}$ obtained on data that ends after a certain fraction of links (panels A and B) and data that ends at a certain time (panels C and D). As in Fig. S7, the correlation value is computed for all users as well as for users whose surprisal value in $\vek{S}^*$ is greater than $10$. Results in panels A and B show that even when one half of links is omitted (note that we omit here the most recent links, not the oldest), correlation between $\vek{S}^*$ and $\vek{S}$ is still around $0.9$. Correlations decreases faster in panels C and D than in panels A and B because the speed at which new links are added increases with time in the studied systems. Setting the end time to $T_{\mathrm{max}}=1,500$ in the Amazon data thus corresponds to using substantially less than $T_{\mathrm{max}}/T_W = 1/2$ of all links.

\end{document}